\documentclass[a4paper]{jpconf}
\usepackage{graphicx}
\begin{document}
\title{SAMPIC: a readout chip for fast timing detectors in particle physics and
medical imaging}

\author{Christophe Royon}

\address{IRFU/Service de Physique des Particules, CEA/Saclay, 91191 Gif-sur-Yvette cedex, France}

\ead{christophe.royon@cern.ch}

\begin{abstract}
We describe the new fast timing readout chip SAMPIC developed in CEA Saclay and
in LAL Orsay (France) as well
as the results of differents tests performed using that chip.
\end{abstract}

\section{Introduction: Timing measurements in particle physics and in medical
imaging}

\begin{figure}[htb]
\centerline{%
\includegraphics[width=10.5cm]{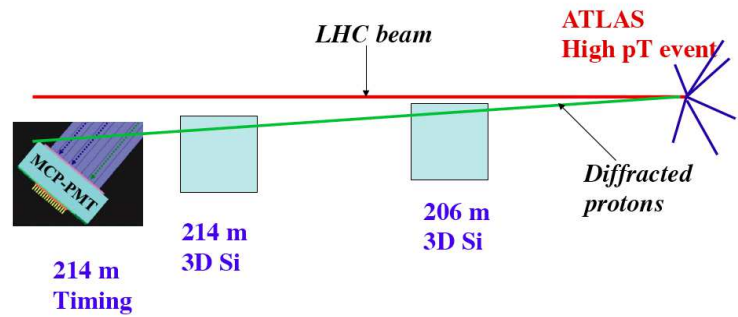}}
\caption{Scheme of the AFP proton detector in ATLAS. The same detector is
implemented on the other side of ATLAS. A similar detector is installed in
CMS/TOTEM.}
\label{Fig1}
\end{figure}

At the Large Hadron Collider (LHC) at CERN, the most energetic proton-proton
collider in the world with a center-of-mass energy of 14 TeV, there are special 
classes of events where protons are found to be intact after collisions. These
events are called ``diffractive" in the case of gluon exchanges. They can originate from
photon exchanges as well. The physics motivation is a better understanding
of diffraction in terms of QCD~\cite{qcd} and the search for beyond standard model physics such
as the existence of extra-dimensions in the universe via
anomalous couplings between photons, $W$ and $Z$ bosons~\cite{anomalous}.
The intact protons scattered at small angles
can be measured in dedicated detectors, hosted in roman pots, located close to 
the beam and far away from the main central ATLAS or CMS detectors. 
In order to measure rare events at the LHC, the
luminosity (or in other words the number of interactions per second) has to be
as large as possible. In order to achieve this goal, the number of interactions
per bunch crossing can be very large, up to 40-70 during the LHC running of
2015-2017 as an example. The projects to measure intact protons at high
luminosity in the ATLAS and CMS/TOTEM experiments are respectively called
AFP (ATLAS Forward Physics) and CT-PPS (CMS/TOTEM-Precision Proton
Spectrometer)~\cite{afpctpps}.
Timing measurements are crucial at the LHC in order to determine if the intact 
protons originate from the main
hard interaction or from secondary ones (pile up). Measuring the proton
time-of-flight with a typical precision of 10 ps allows constraining the 
protons to originate from the main interaction point of the event (hard
interaction) with a precision of about 2.1 mm. For a pile up of 40 (which means
about 40 interactions occuring in the same bunch crossing at the LHC), such a
precision on time-of-flight measurements 
leads to a reduction in background of a factor of about 40~\cite{matthias}.

Timing measurements have also many applications in drone technology and in
medical imaging as an example. The ``holy grail" of medical imaging would be a
PET detector with a 10 ps timing precision. With such an apparatus, image
reconstruction is no longer necessary (the analysis can be performed online)
since many fake coincidences can be suppressed, only attenuation corrections are
needed, and real time image formation can be performed. 

In order to achieve a 10 ps precision, many steps are needed going from the
detector to the electronics and the readout software. In this article, 
we will concentrate on the achievement concerning the picosecond timing 
electronics that is currently being done in IRFU/SEDI Saclay and in
LAL-Orsay~\cite{Eric, Dominique}.

\section{SAMPIC: SAMpler for PICosecond time pick-off}

\begin{figure}[htb]
\centerline{%
\includegraphics[width=13.5cm]{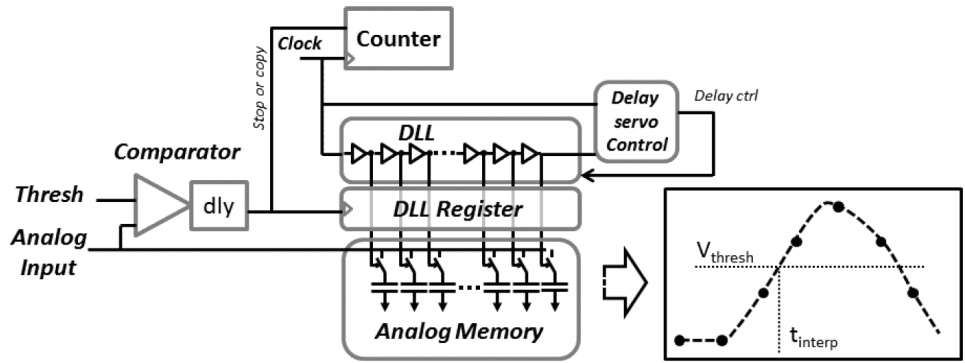}}
\caption{Scheme of the SAMPIC chip.}
\label{Fig2}
\end{figure}

\begin{figure}[htb]
\centerline{%
\includegraphics[width=10.5cm]{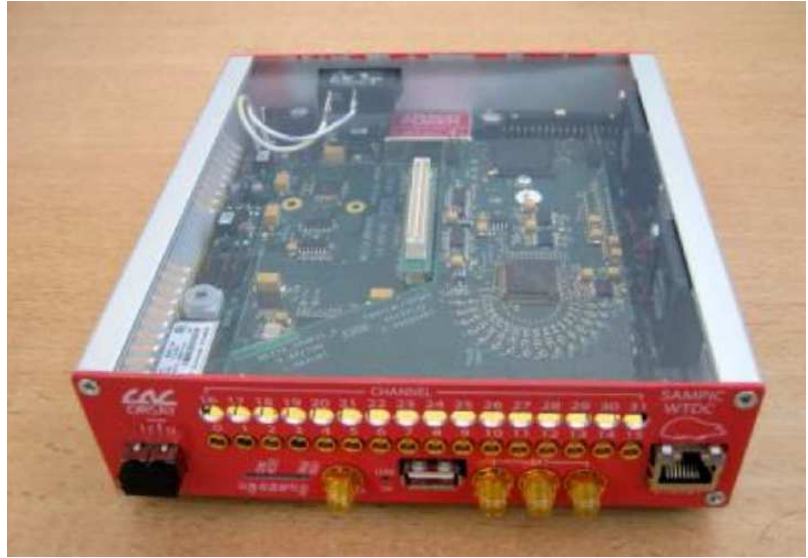}}
\caption{Picture of the SAMPIC chip and its acquisition board.}
\label{Fig3}
\end{figure}

Before SAMPIC, the most performant Time to Digit Converters (TDCs) used digital
counters and Delay Line Loops (DLLs). The timing resolution is limited by the
DLL step and with most advanced Application Specific Integrated Circuit 
(ASICs), one gets a resolution of about 20 ps (new
developments at CERN target 5 ps). The inconvenient is that a TDC needs a
digital input signal: the analog input signal has to be transformed into a
digital one with a discriminator which means that the timing resolution will be
given by the quadratic sum of the discriminator and the TDC timing resolutions,
thus leading to worse timing resolutions.

A new approach had to be developed using the principle of a wafeform based TDC.
The idea is to acquire the full waveform shape of a detector signal in an
ASIC dedicated to picosecond timing
measurements. The input signal range has to be between 0.1 and 1. V with a fast
rising time up to 1.ns in order to get the best possible performance of SAMPIC.
The present version of the chip holds 16 channels (50 $\Omega$ terminated) with
independent dead time. The possible trigger modes are either self triggered or
triggered externally. Each channel includes an analog memory (64 cells) and
recording is triggered by a discriminator. A Gray counter associated to DLLs
allows assigning a time to the different samples and an ADC provides the
conversion into a digital signal. 

Three timing measurements with different precision are performed in SAMPIC.
The time stamp Gray counter has a 6 ns step (it samples the reference clock),
the DLL 150 ps (it defines a region of interest) and the waveform shape a few ps
RMS after interpolation between the acquired points (they are acquired on a 64
step analog memory).

As we already mentioned, SAMPIC acquires the full waveform shape of a detector
signal. The discriminator is used only for triggering, not for timing, and thus
there is no jitter originating from the discriminator. All the information
concerning the signal is kept in SAMPIC, and it is possible to use offline
signal processing algorithms in order to improve the timing resolution. It can
also be used to obtain other signal characteristics such as the deposited
charge. In the present version, SAMPIC suffers an important dead time per
channel due to the ADC conversion of about 1 $\mu$s. It will be reduced by
about one order a magnitude in the next version of SAMPIC, using in particular
the so called ``ping-pong" method and analog buffering. Two SAMPIC chips can be hosted in a
mezzanine board developed in LAL, Orsay, leading to a 32-channel system. 
The input into SAMPIC is sent via MCX connectors. SAMPIC can
be read out using an USB-Ethernet-Optic fiber readout is also
provided. A 5 V voltage power supply is the only element needed to run the
mezzanine board and the readout software runs on Windows or Linux.
A scheme of SAMPIC is given in
Fig.~\ref{Fig2} and a picture of SAMPIC together with its acquisition board in
Fig.~\ref{Fig3}.

SAMPIC is quite cheap (about 10 Euros per channel) with respect to a few 1000s
Euros for previous technology, which means that it can be used in large scale
detectors such as PET for medical applications.

\begin{figure}[htb]
\centerline{%
\includegraphics[width=13.5cm]{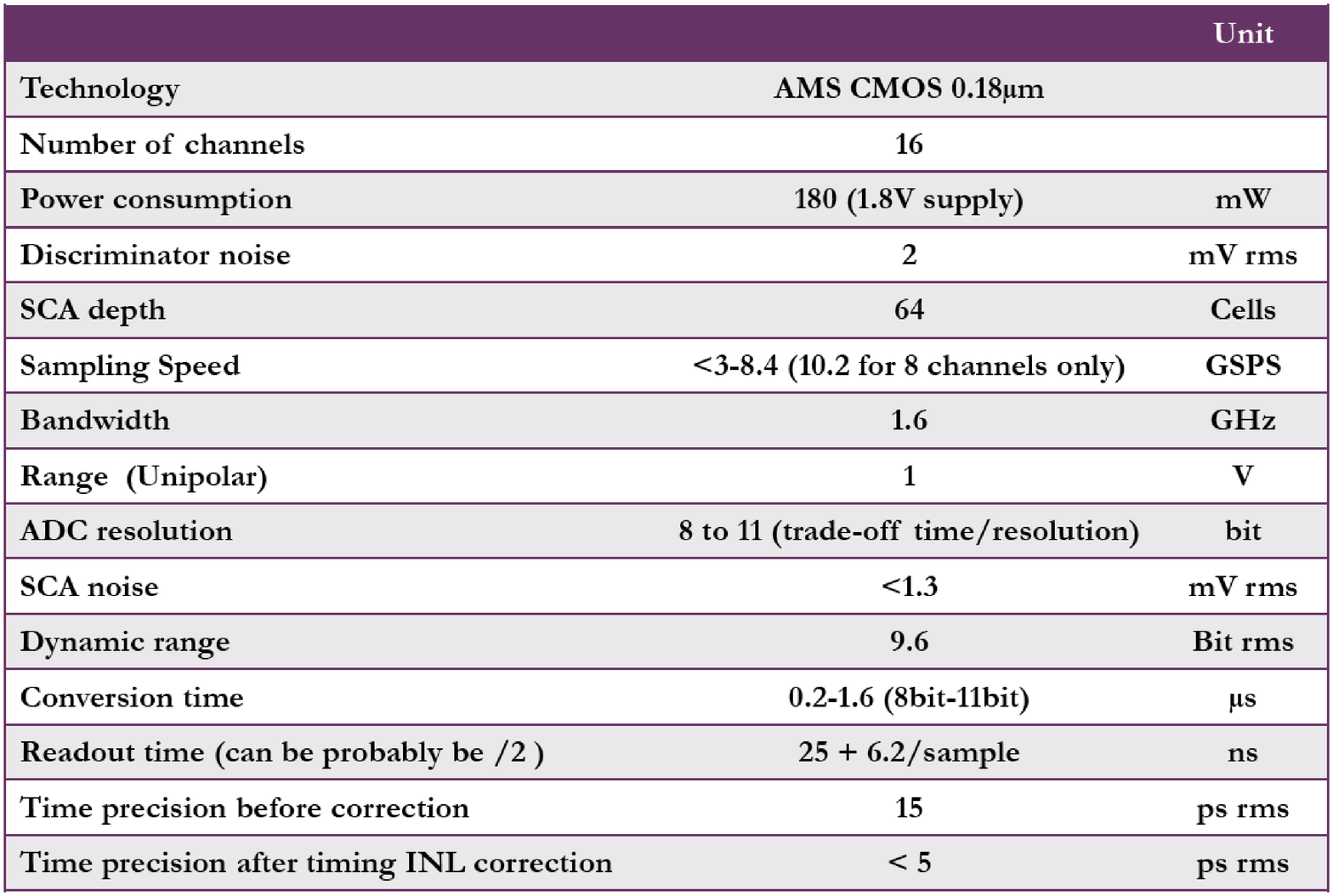}}
\caption{Parameters of the SAMPIC chip.}
\label{Figa}
\end{figure}

As a reference, a table giving the parameters of the SAMPIC chip is given in
Fig.~\ref{Figa}.

\section{SAMPIC performance}

\subsection{Electronics tests}
In this section, we describe the SAMPIC performance obtained from 
pure electronics tests. The maximum signal size is about 1.V, and after
corrections, the average noise is quite low, of about 1 mV RMS (the noisiest cells being 1.5 mV
RMS), which means a dynamic 10 bits RMS. 

The SAMPIC cross talk was measured by sending a signal of 800 mV with a 300 ps
rise time on one channel and reading out the neighbouring channels. The cross
talk was found to be less than 1\%. The quality of sampling was tested using a
sinus wave signal, and the signal was perfectly reproduced without corrections
at a sampling frequency of 10 Gigasamples per second. The sampling speed in SAMPIC is possible between 3
and 8.2 Gigasamples per second on 16 channels (up to 10 Gigasamples per second
for 8 channels).

\begin{figure}[htb]
\centerline{%
\includegraphics[width=10.5cm]{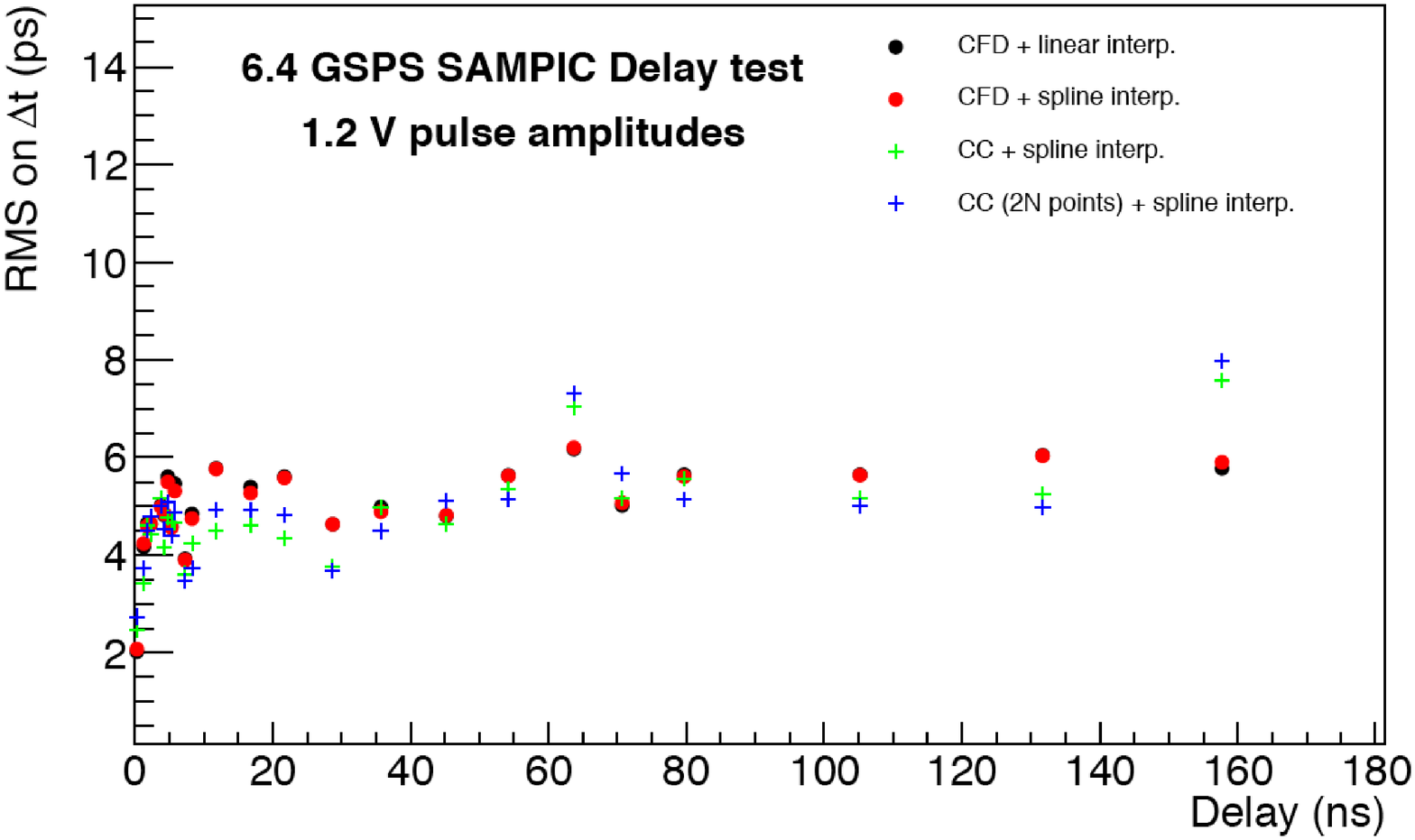}}
\caption{RMS on the time difference between two signals, one being delayed with
respect to the other using two offline algorithms (Constant Fraction
Discriminator (CFD) and cross correlation (CC)) using a linear or a spline
interpolation.}
\label{Fig4}
\end{figure}

The timing resolution was studied by using two different channels of SAMPIC.
The same signal was sent on both channels, one being delayed compared to the
other one using a delay box or longer cables. The pulse had an amplitude of about 
1.2 V, and we used the 6.4 Gigasamples per second configuration. 
The RMS of the time difference between the two
signals as a function of delay is given in Fig.~\ref{Fig4} using two different
offline algorithms to reconstruct the time difference (CDF as constant fraction
discriminator and CC as cross correlation using a linear or a spline
interpolation between the different points measured by SAMPIC). The time
resolution is quite flat as a function of the delay between the two signals and
is about 5 ps, leading to a time resolution per channel of about 4 ps.

\begin{figure}[htb]
\centerline{%
\includegraphics[width=10.5cm]{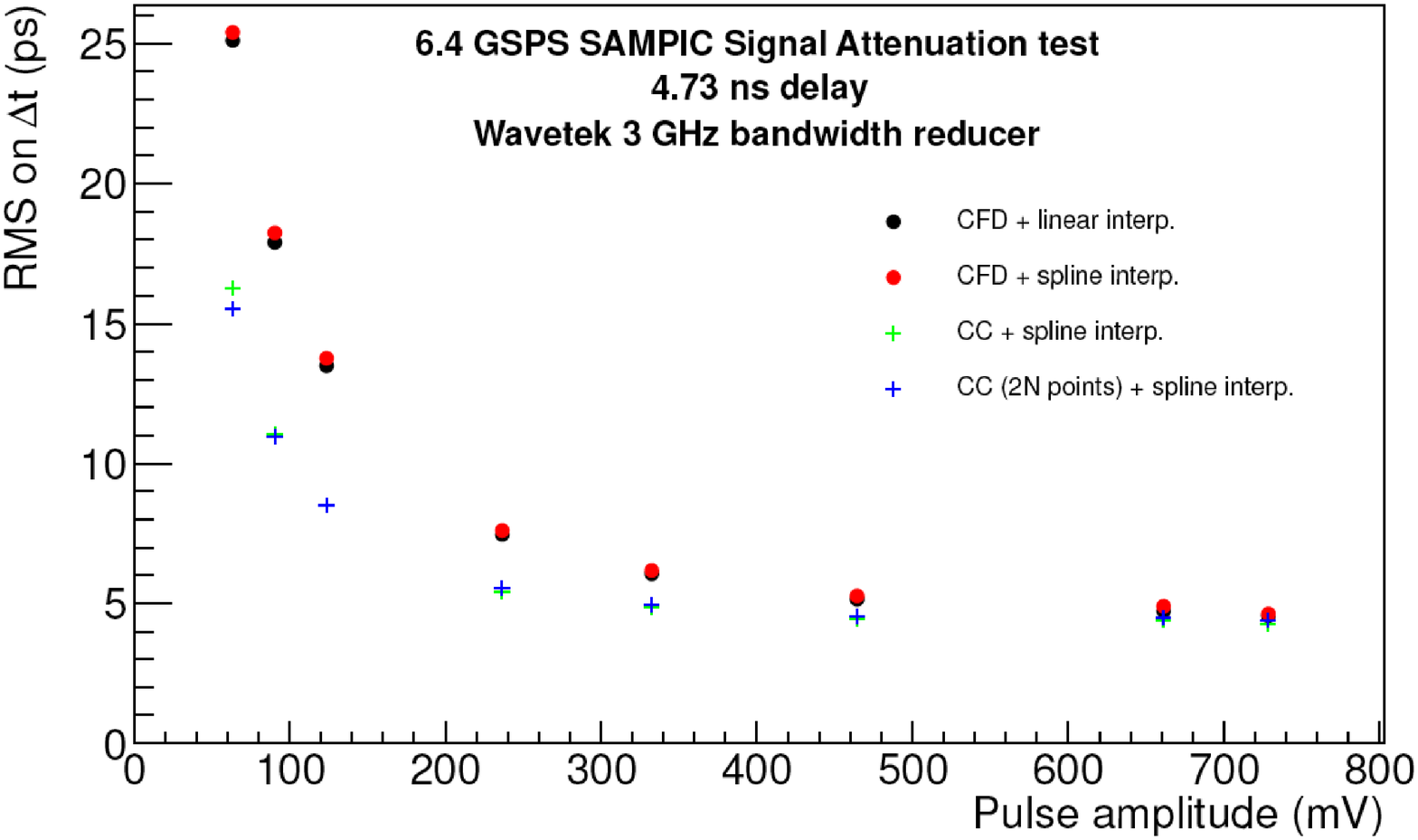}}
\caption{RMS on the time difference between two signals as a function of
the signal amplitude using two offline algorithms (Constant Fraction
Discriminator (CFD) and cross correlation (CC)) using a linear or a spline
interpolation.}
\label{Fig5}
\end{figure}

A similar study of the timing resolution versus the signal amplitude is shown in
Fig.~\ref{Fig5}. The signal has to be above 450 mV in order to obtain the best
timing resolution possible of about 4 ps.

\subsection{Timing resolution using detectors}
\begin{figure}[htb]
\centerline{%
\includegraphics[width=10.5cm]{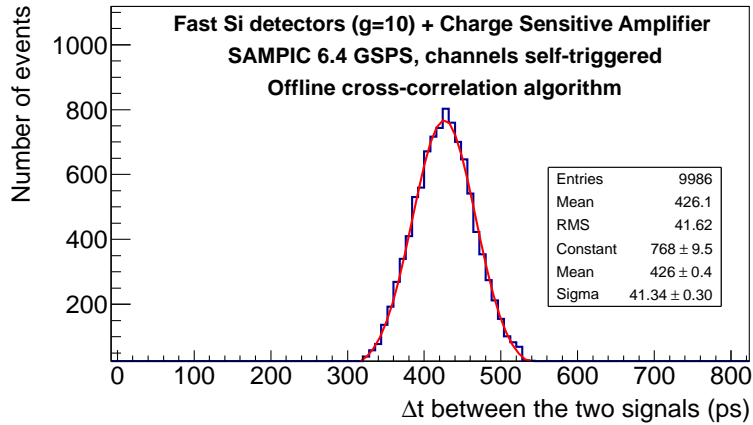}}
\caption{Time difference between two SAMPIC channels reading out Si detectors,
a laser signal splitted in two going through the two Si detectors.}
\label{Fig6}
\end{figure}

The second series of tests was performed by plugging SAMPIC into a real
detector. We used a laser signal splitted in two, and going through two fast Si
detectors~\cite{nicolo}. The time difference between the two channels was measured using
SAMPIC. The result is shown in Fig.~\ref{Fig6} using the offline cross
correlation algorithm. The time resolution is about 30 ps. It is of course
dominated by the fast Si detectors, the resolution of SAMPIC being of the order
of 4 ps. Additional studies are being performed in beam tests using diamond
detectors leading to a time resolution of 80 to 90 ps~\cite{TOTEM}.

\section{Conclusion}

A self triggered timing chip demonstrator has been designed and characterised
with 1.6 GHz bandwidth, up to 10 Gigasamples per second, low noise and of the
order of 4 ps timing resolution. The chip is now ready and can be used for
tests. Tests already started within the AFP, CT-PPS and CMS/TOTEM projects using
quartz, diamond and Si detectors. Work is still going on in order to improve the chip
concerning the DAQ system optimisation (firmware and software) and the
improvement of the dead time using the ``ping-pong" method. SAMPIC can now be
used in many applications for tests in addition to particle physics for instance
in medical imaging, drones, in detectors including many channels due to the low
cost per channel.  

\section*{Acknowledgments} These results originate from a useful collaboration
with D. Breton, V. de Caqueray, N. Cartiglia, E. Delagnes, H. Grabas, J. Maalmi, N. Minafra,
M. Saimpert.

\end{document}